\documentclass[fleqn,10pt]{wlscirep}
\usepackage[utf8]{inputenc}
\usepackage[T1]{fontenc}
\usepackage{amsmath, nccmath}
\usepackage{geometry}
\title{Multiplicity dependence of the freezeout parameters in high energy hadron-hadron collisions}

\author[1,*]{Muhammad Ajaz} 
\author[1]{Majid Shehzad}
\author[2]{Muhammad Waqas}
\author[3,*]{Haifa I. Alrebdi} 
\author[4]{Momhammad Ayaz Ahmad} 
\author[4]{Antalov Jagnandan} 
\author[4]{Shawn Jagnandan} 
\author[1]{Murad Badshah}
\author[5] {Jalal Hasan Baker} 
\author[6] {Abdul Mosawir Quraishi} 
\affil[1]{Department of Physics, Abdul Wali Khan University Mardan, Mardan 23200, Pakistan}
\affil[2] {School of Mathematics, Physics and Optoelectronic Engineering, Hubei University of Automotive Technology, Shiyan 442002, China}
\affil[3]{Department of Physics, College of Science, Princess Nourah bint Abdulrahman University, P.O. Box 84428,
Riyadh 11671, Saudi Arabia}
\affil[4]{Department of Mathematics, Physics and Statistics, Faculty of Natural Sciences, University of Guyana, 101110 Georgetown, Guyanan, South America}
\affil[5]{Department of  Physics, Faculty of Science, University of Tabuk, Tabuk, Kingdom of Saudi Arabia}
\affil[6]{Department of Electrical Engineering, College of Engineering, Qassim University, Unaizah, Saudi Arabia}

\affil[*]{ajaz@awkum.edu.pk (M.Ajaz), hialrebdi@pnu.edu.sa (H.I.Alrebdi)}


\begin{abstract}
We examined the transverse momentum ($p_T$) spectra of various identified particles, encompassing both light-flavored and strange hadrons ($\pi^+ + \pi^-$, $K^+ + K^-$, $p + \Bar{p}$, $\phi$, $K_s^0$, $\Lambda + \Bar{\Lambda}$, $\Xi^- + {\Bar{\Xi}}^+$, $\Omega^- + {\Bar{\Omega}}^+$), across different multiplicity classes in proton-proton collisions (p-p) at a center-of-mass energy of $\sqrt{s}$ = 7 TeV. Utilizing the Tsallis and Hagedorn models, parameters relevant to the bulk properties of nuclear matter were extracted. Both models exhibit good agreement with experimental data. In our analyses, we observed a consistent decrease in the effective temperature (T) for the Tsallis model and the kinetic or thermal freeze-out temperature ($T_0$) for the Hagedorn model, as we transition from higher multiplicity (class-I) to lower multiplicity (class-X). This trend is attributed to the diminished energy transfer in higher multiplicity classes. Additionally, the transverse flow velocity ($\beta_T$) experiences a decline from class-I to class-X. The normalization constant which represents the multiplicity of produced particles is observed to decrease as we move towards higher multiplicity classes. While the effective and kinetic freeze-out temperatures, as well as the transverse flow velocity, show a mild dependency on multiplicity for lighter particles, this relationship becomes more pronounced for heavier particles. The multiplicity parameter for heavier particles is noted to be smaller in comparison to lighter particles, indicating a greater abundance of lighter hadrons compared to the heavier ones. Various particle species are observed to undergo decoupling from the fireball at distinct temperatures: lighter particles exhibit lower temperatures, while heavier ones show higher temperatures, thereby supporting the concept of multiple freeze-out scenarios. Moreover, we identified a positive correlation between the kinetic freeze-out temperature and transverse flow velocity, a scenario where particles experience stronger collective motion at higher freeze-out temperature. The reason for this positive correlation is that as the multiplicity increases, more energy is transferred into the system. This heightened energy causes greater excitation and pressure within the system, leading to a quick expansion.   
\end{abstract}
\begin{document}

\flushbottom
\maketitle
%
%
\thispagestyle{empty}

\section*{Introduction}

The collision of relativistic heavy nuclei results in a highly unstable, transient state of matter known as Quark Gluon Plasma (QGP), existing for a brief duration. The name Quark Gluon Plasma refers to this state of matter  (QGP)\cite{busza2018heavy, das2016propagation, das2017effect, gelis2004photon} which is a strongly interacting matter where quarks and gluons are asymptotically free which is a peculiar behaviour. The temperature and energy density of QGP is as high as those of the early universe; due to this resemblance, QGP is the hot field for research in high-energy physics, and many collaborations around the globe are working on QGP to explore the conditions of the early universe. Though the existence of QGP has been confirmed when heavy nuclei like Pb-Pb or Au-Au collide, many signatures of the QGP have also been observed in p-p collisions. Mainly, these collisions provide baselines for heavy-ion collisions. These signatures of QGP may include strangeness enhancement, $J/\psi$ melting or suppression, jet quenching, etc. \cite{5, 6, 7, 8}.

We used p-p collisions as references to study the collision mechanism of heavy nuclei at the colliders. Recently results obtained from p-p and proton-lead (p-Pb) collisions have shown similar trends to that of Pb-Pb collisions. For instance, spectra of $p_T$ as a function of transverse mass $m_T$ have shown the same excitation function with charged hadron multiplicity as shown by the larger colliding systems. The analyses of the $p_T$ distribution in p-p collisions show a clear hardening of the $p_T$ spectra from multiplicity class-X to class-I, where in class-X the average charged particle densities are lowest while in class-I the charged particle densities are highest, which is the same trend observed in collisions of larger systems \cite{identified}. Additionally, it has been found that, in contrast to pions, the event charged-particle multiplicity increases with the integrated yields of strange and multi-strange hadrons \cite{strange}. In Pb-Pb collisions, this strangeness enhancement has already been noted.            

QGP behaves like a hot fluid that expands and cools down; during its evolution, it passes through many stages, and each stage has its corresponding temperature. Chemical freeze-out temperature corresponds to the chemical freeze-out stage where inelastic collisions among the constituents of the fireball vanish which results in the stoppage of new particle production. The reason behind this is that when the fireball expands due to pressure gradient then at a certain point the inter-particle distances become so large that the constituents of the fireball can not interact inelastically \cite{9, 10, 11}. The thermal freeze-out stage corresponds to the kinetic or thermal freeze-out temperature, where elastic collisions between the generated particles also come to an end \cite{12, 13, 14}. After the thermal freeze-out stage, the particles move outward, where they may be detected by the detectors, and the particles at the detectors have the same energy and momentum spectra or distribution as they have at the thermal freeze-out stage. This implies very accurate information about the thermal or kinetic freeze-out stage by studying the $p_T$ spectra of the outgoing particles. The information about the freeze-out stage may include the thermal freeze-out temperature, the radial or transverse flow velocity of the outgoing particles, the volume of the created system (V) after the collision, the number of particles created or the multiplicity parameter ($N_0$), etc.   

The detectors cannot measure the above-stated freeze-out parameters; therefore, many hydrodynamical and statistical models are used to extract the values of these valuable parameters, which help in understanding the true nature of QGP. Even after the interactions between particles stop during the fireball's evolution, the particles still occupy space based on a statistical distribution \cite{15}. The models used for the analysis in the present work and those which are commonly used will now be briefly discussed here. It is very popular and convenient to use non-extensive statistical distributions to analyze the $p_T$ spectra in high-energy collisions. The literature has employed a variety of Tsallis distributions to effectively explain the $p_T$ spectra of produced particles in p-p interactions at the RHIC and LHC energies \cite{16, 17, 18, 19, 20,21}. The best thing about using the Tsallis distribution is that it only depends on three parameters that help in fitting the experimental data; the first parameter is the effective temperature (T) which includes both thermal and flow effects of the fireball, the second one is the non-extensive variable (q) which measures the deviation from the extensive Boltzmann-Gibbs statistics, and the third one is the fitting or normalization constant ($N_0$). These three parameters can be incorporated to calculate the system's volume and initial conditions like initial temperature etc \cite{22}. To analyze the $p_T$ distributions of the outgoing particles in high energy nuclei collisions at the LHC and RHIC, the various flow models are incorporated into Tsallis statistics. To extract radial flow velocity and kinetic freeze-out temperature, mainly the Hagedorn formula with the embedded transverse flow \cite{23, 24, 25, 26}, the Blast-Wave model with Tsallis statistics (TBW model) \cite{27, 28}, Blast-Wave model with Boltzmann Gibbs statistics (BGBW model) \cite{29, 30} and the Tsallis distribution with flow effect or the improved Tsallis distribution  \cite{31, 32, 33} have been used in the literature.

The rest of the paper is structured as follows: the 'Methodology' section elucidates the statistical models utilized in our analyses. Subsequently, the 'Results and Discussion' section presents our findings in detail and provides accompanying explanations. Finally, the 'Conclusion' section encapsulates the outcomes and deductions drawn from our research endeavors.

\section*{Methodology}
Experimental $p_T$ distributions for identified light and strange particles in $\sqrt{s}$ = 7 TeV p-p collisions were sourced from \cite{identified} and \cite{strange}, respectively. 
The $p_T$ distributions of particles have a lot of information about the freeze-out stage of QGP therefore these distributions are very significant for the extraction of these pieces of information (parameters). Various statistical and hydrodynamical models are being used to extract these parameters from $p_T$ distributions as described above in the 'Introduction' section. In the present paper, two models, namely the Tsallis model and the Hagedorn model, have been used. The classical exponential function called Boltzmann Gibbs function, given in Eq. 1, is found to be only suitable for low $p_T$ regions. For high $p_T$ regions a power law distribution is found to be more appropriate. Tsallis function, given in Eq. 2, is one of such power law distribution functions which covers a wide range of $p_T$.
\begin{align}
f(E) \propto exp (-\frac{E-\mu}{T})
\end{align}
\cite{}\begin{align}
\frac{d^2N}{\ N_{evt}\ dp_T\ dy}=2\ \pi\ C\ p_T \bigg[1+\frac{(q-1)}{T}m_T\bigg]^{-\frac{1}{(q-1)}}
\end{align}
In Equation 2, {$N$}, $N_{evt}$ and $p_T$ represent the number of particles produced in the collision, the number of participating particles taking part in the collision and the transverse momentum respectively \cite{33a}, T signifies the effective temperature, encompassing both the thermal and flow effects within the evolving system. The non-extensivity parameter is denoted as q, and C stands for the fitting constant, which is directly related to the size or volume of the created system as given by the equation $C=\frac{g V}{(2 \pi)^3}$, where V denotes the system's volume and g is the degeneracy factor with varying values for different particles \cite{34, 34a, 37a}. The transverse mass of the outgoing particle is represented by $m_T$, calculated using the formula $m_T= \sqrt{{p_T}^2- {m_0}^2}$, where $m_0$ corresponds to the rest mass of the produced particle. The Tsallis distribution function is utilized in the literature in a variety of ways, the thermodynamically most consistent form is given in Eq. 3.
\begin{align}
\frac{d^2N}{\ N_{evt}\ dp_T\ dy}=2\ \pi\ C\ p_T\ m_T \bigg[1+\frac{(q-1)}{T}m_T\bigg]^{-\frac{q}{(q-1)}}
\end{align}
It is important to note that the Tsallis model presented in Eq. 2 or Eq. 3 provides insights into the parameter T, but doesn't offer information about $T_0$ and $\beta_T$. In order to extract these two parameters, we employed the Hagedorn model, which incorporates $\beta_T$. The functional expression for this model is outlined in the simplified form provided by Eq. 4.
\begin{equation}
\frac{d^2N}{\ N_{evt}\ dp_T\ dy}=2\ \pi\ C \ p_T \bigg[1+\frac{m_T}{nT_0}\bigg]^{-n}
\end{equation}
Where $T_0$ is the thermal freeze-out temperature and n is the entropy parameter. To have a contribution from the transverse flow velocity ($\beta_T$), replace $m_T$ by $<\gamma_T> (m_T - P_T <\beta_T>)$ in Eq. 4 \cite{36a}, where $<\gamma_T>$ is a Lorentz factor, Eq. 5 is the simplest form of Hagedorn model with the embedded $\beta_T$.
\begin{align}
\frac{d^2N}{\ N_{evt}\ dp_T\ dy}=2\ \pi\ C \ p_T \bigg[1+\frac{<\gamma_T> (m_T - P_T <\beta_T>)}{nT_0}\bigg]^{-n}
\end{align}
It is important to mention that in our current analysis, we used the minimum $\chi^2$ method to fit the theoretical model functions on the experimental transverse momentum spectra of particles. The minimum $\chi^2$ method considers combined statistical and systematic errors added in quadrature.

\section*{Results and discussion} 
Figures \ref{fig:01}(a)-\ref{fig:01}(h) show the transverse momentum distribution, $\frac{d^2N}{\ N_{evt}\ dp_T\ dy}$, for various light-flavored identified and strange hadrons in 7 TeV p-p collisions categorized into various multiplicity classes. Each plot represents the $p_T$ distribution for a specific particle type, including $\pi^++\pi^-$, $K^++K^-$, $K_s^0$, $p+\bar p$, $\phi$, $\Lambda+\bar\Lambda$, $\Xi^-+\bar\Xi^+$, and $\Omega^-+\bar\Omega^+$. Within each plot, different colors correspond to different multiplicity classes in the experimental data. While distinct symbols across Figures \ref{fig:01}(a)-\ref{fig:01}(h) are used for different particle species. Overlaid on these data points, solid and dotted curves depict the fit outcomes derived using the Tsallis distribution (Eq. 3) and the Modified Hagedorn model with incorporated flow (Eq. 5), respectively. In some cases, scaling factors are applied to certain spectra to prevent curve and data point overlap within a single plot. These scaling factors are specified alongside each multiplicity class at the upper portion of each plot. During the fitting process, efforts are made to minimize the $\chi^2$ value for each fit, aiming to achieve high-quality fits and thus accurate parameter extraction. The extracted parameters by the Tsallis and Hagedorn models are tabulated in tables \ref{tab:1} and \ref{tab:2} respectively. The constant "$C$" as presented in Equations 2 to 5, serves as a normalization constant ensuring that the integral of the functions in these equations evaluates to unity. On the other hand, "$N_0$" is another normalization constant, appeared in tables \ref{tab:1} and \ref{tab:2}, used for comparing the function or model against experimental data. Despite the possibility of absorbing $C$ into $N_0$, both constants retain distinct purposes. The presence of both $C$ and $N_0$ allows for precise descriptions within the context of the study.

\begin{figure}[p!]
\centering
\includegraphics[width=8.2cm]{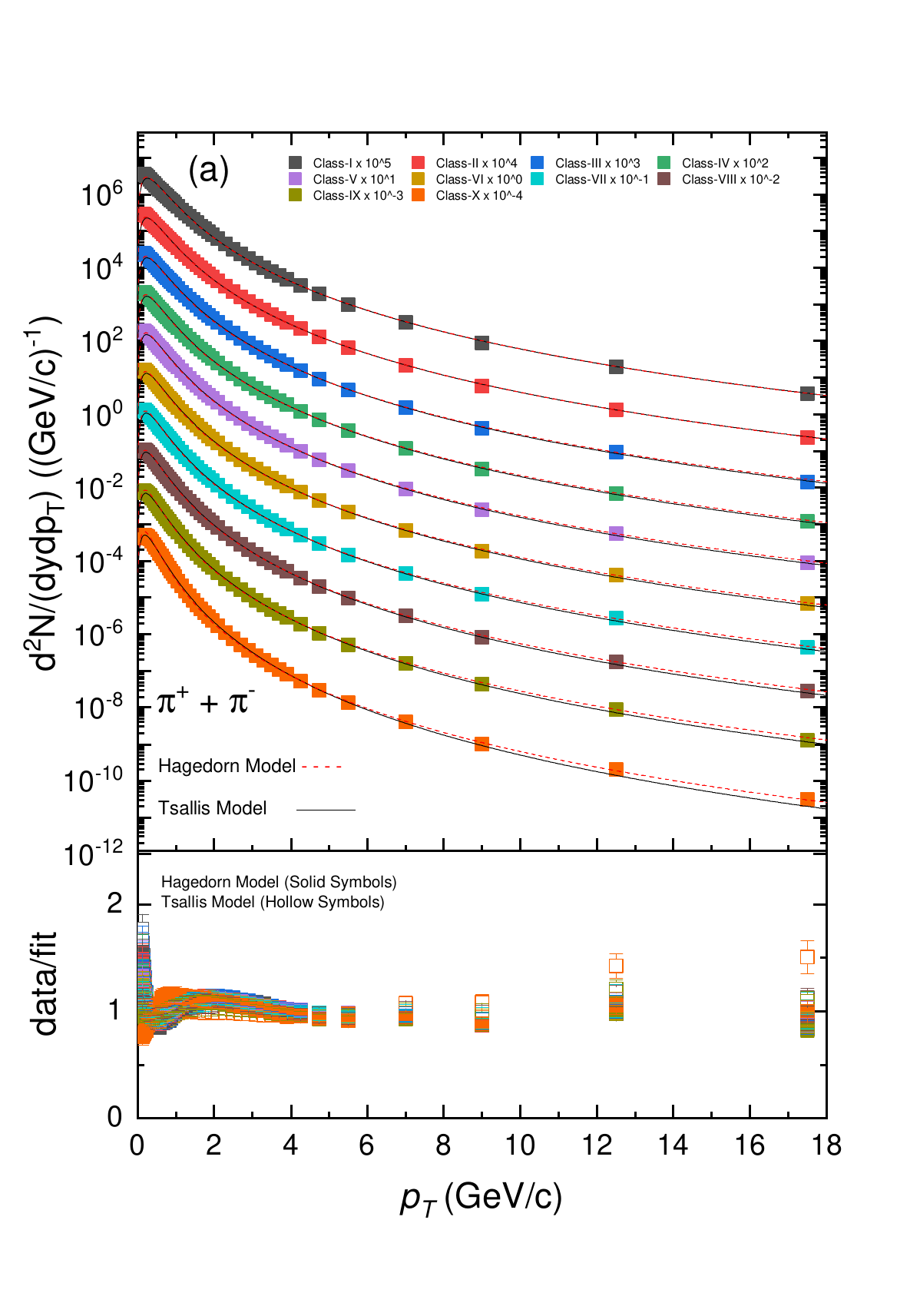}
\vspace{-0.7cm}
\hspace{-0.6cm}
\includegraphics[width=9cm]{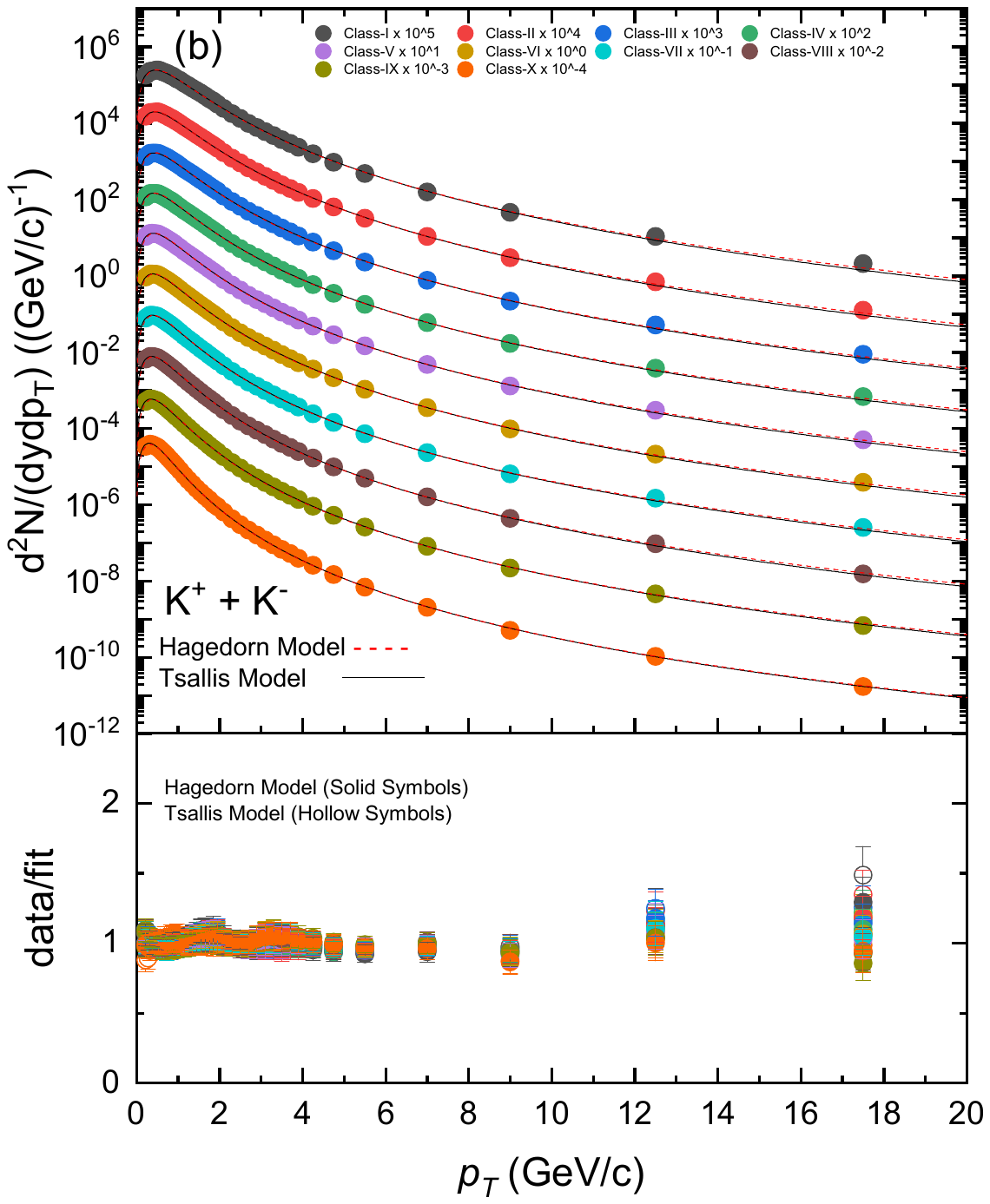}
\includegraphics[width=8.2cm]{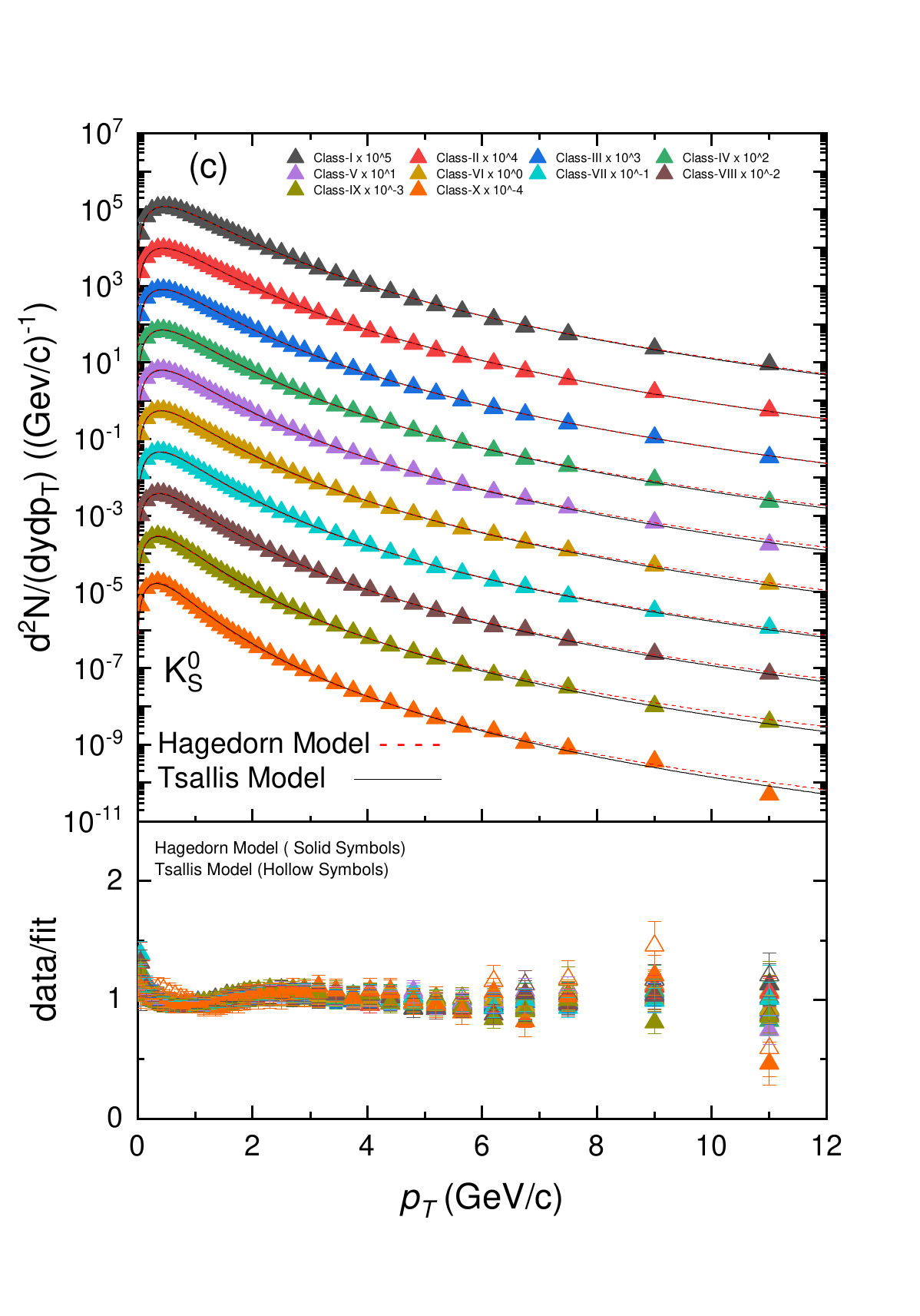}
\hspace{-0.6cm}
\includegraphics[width=9cm]{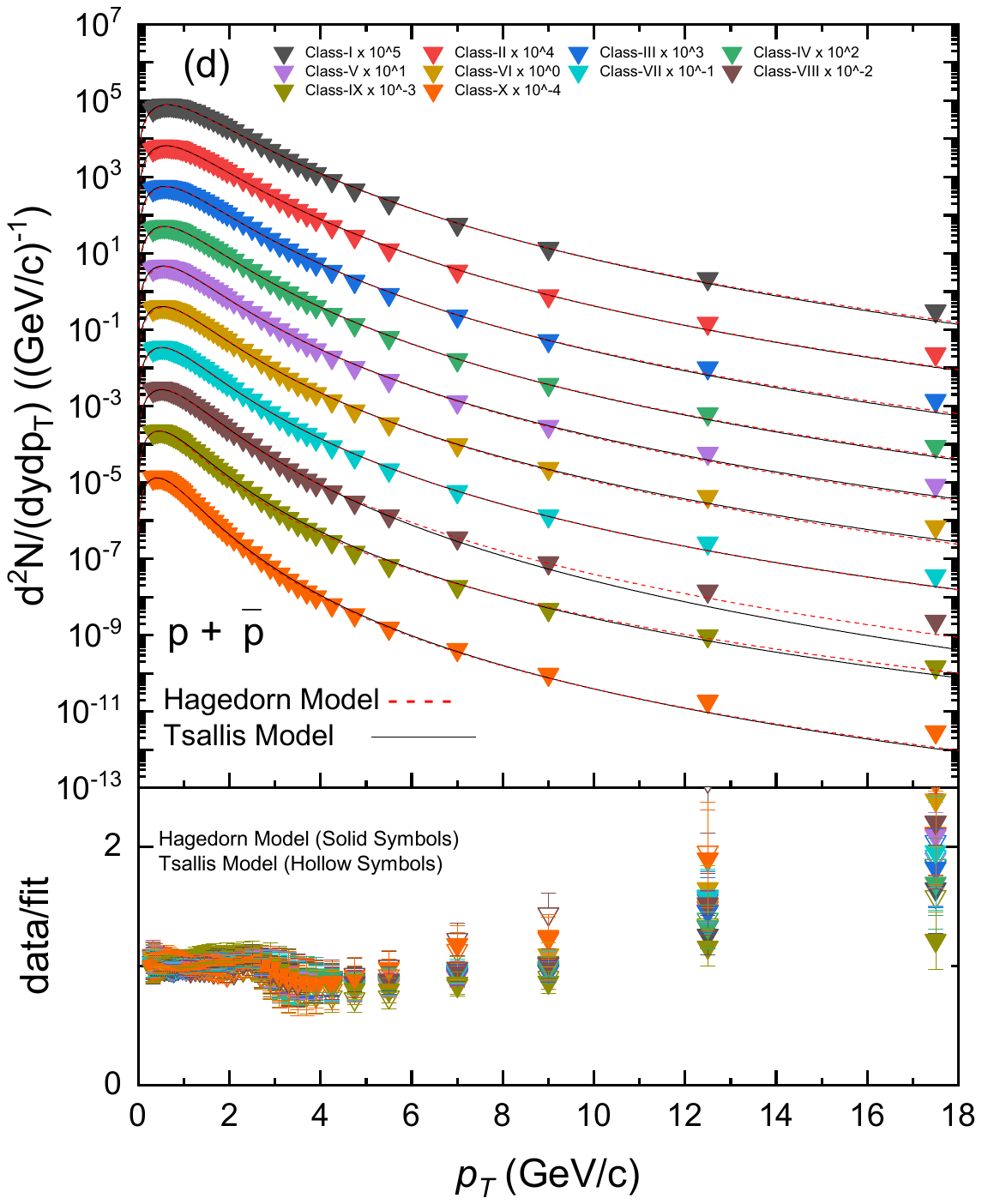}
\end{figure}

\begin{figure}[p!]
\centering
\includegraphics[width=8.2cm]{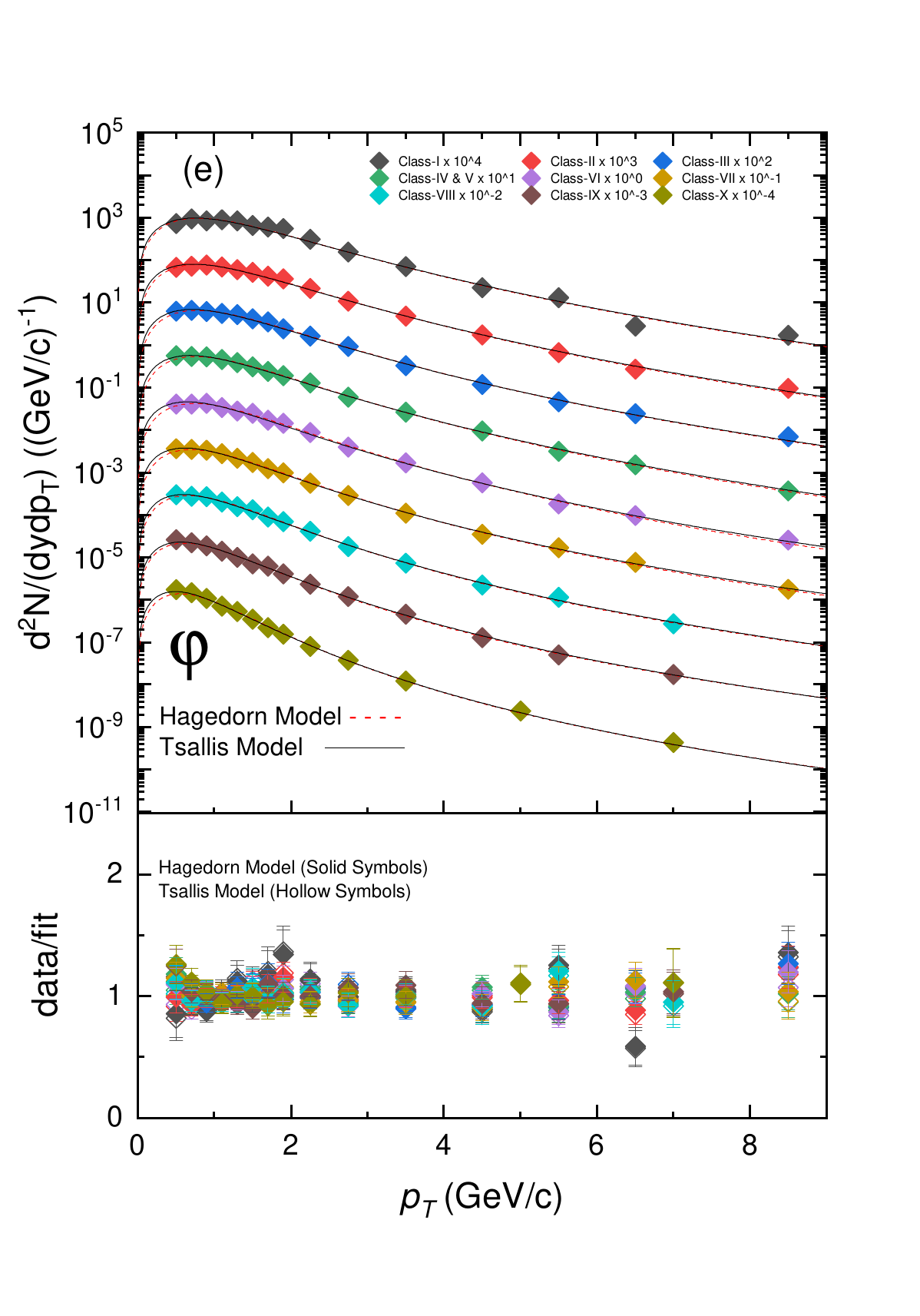}
\vspace{-0.7cm}
\hspace{-0.6cm}
\includegraphics[width=9cm]{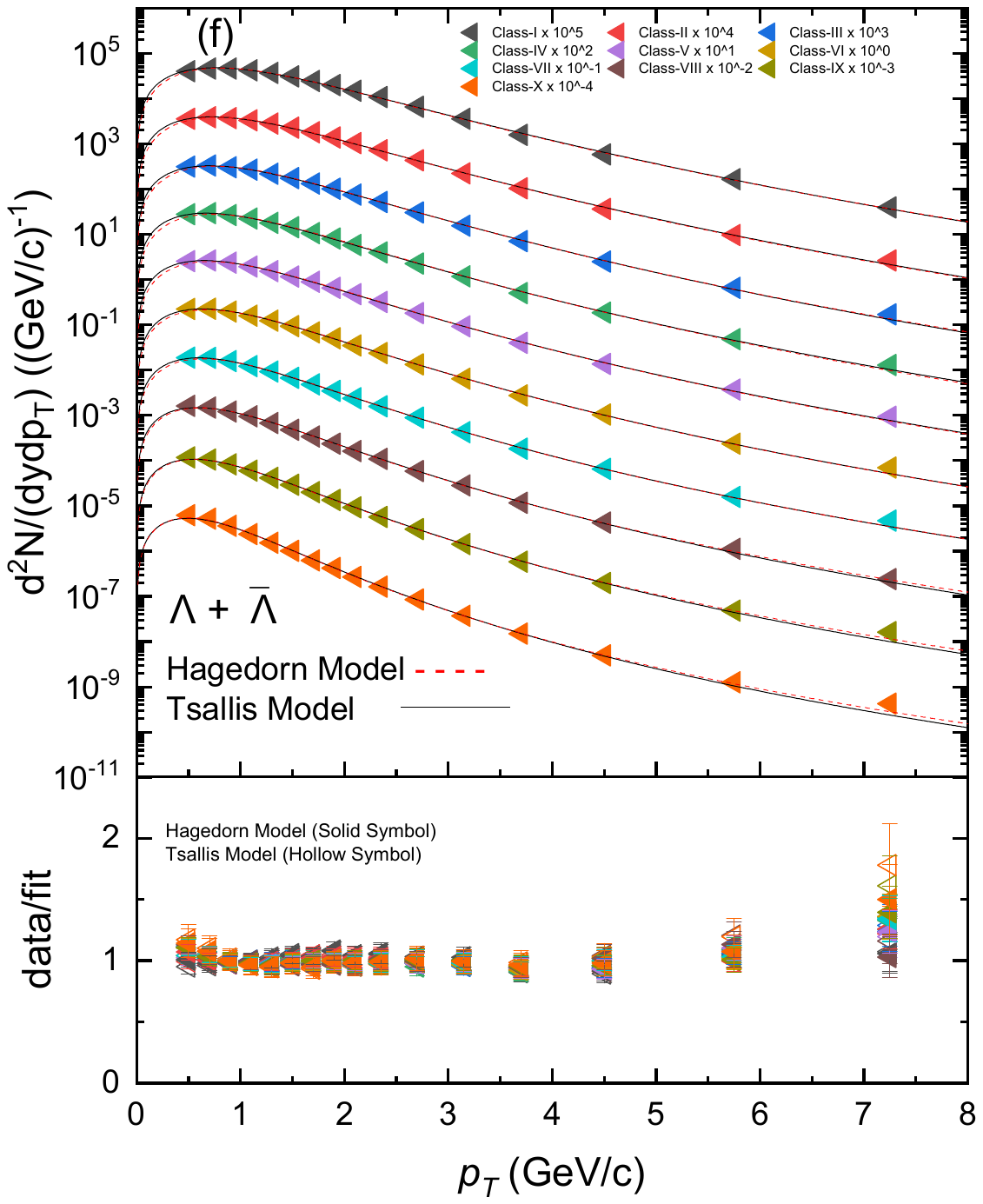}
\includegraphics[width=8.2cm]{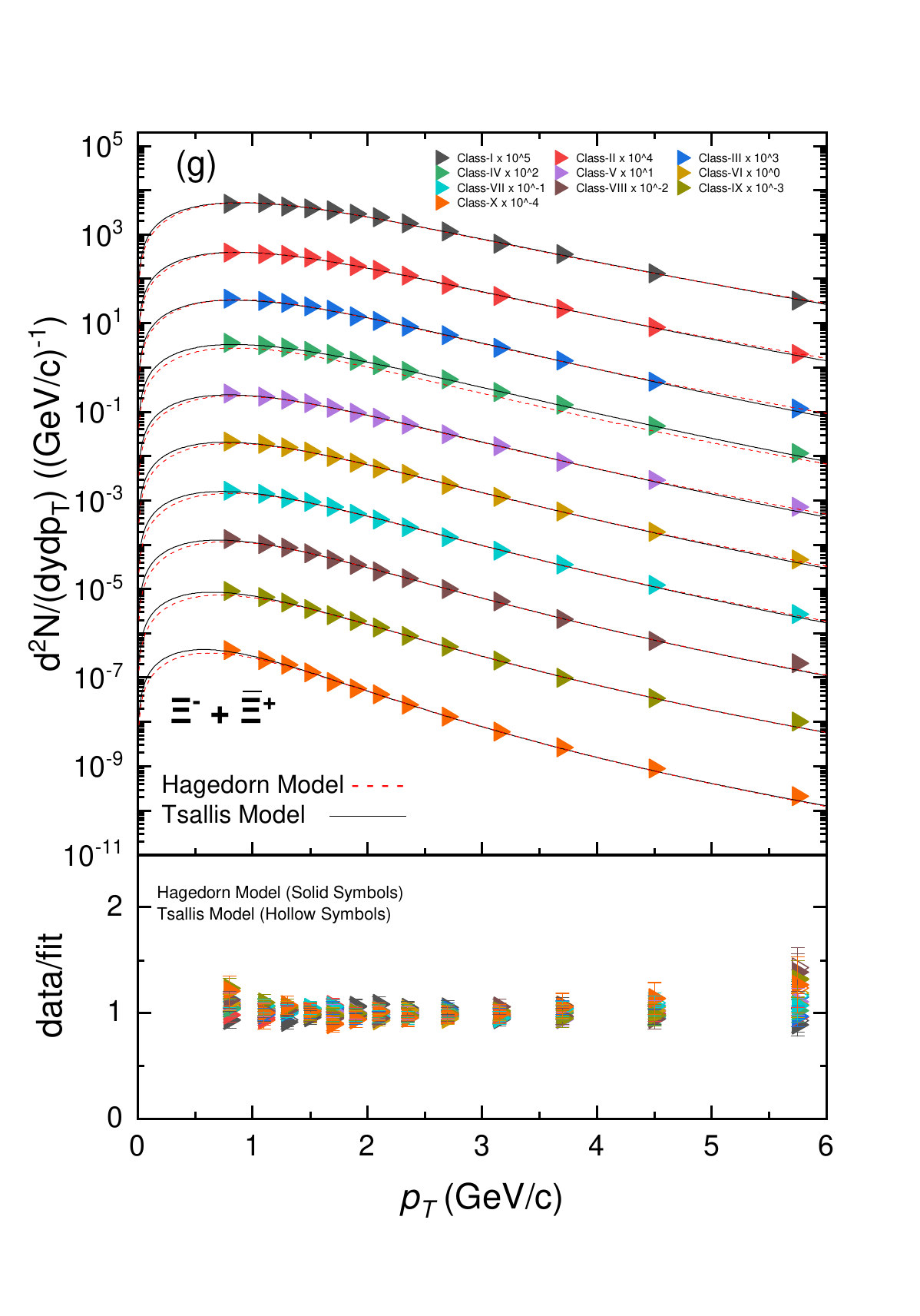}
\hspace{-0.6cm}
\includegraphics[width=9cm]{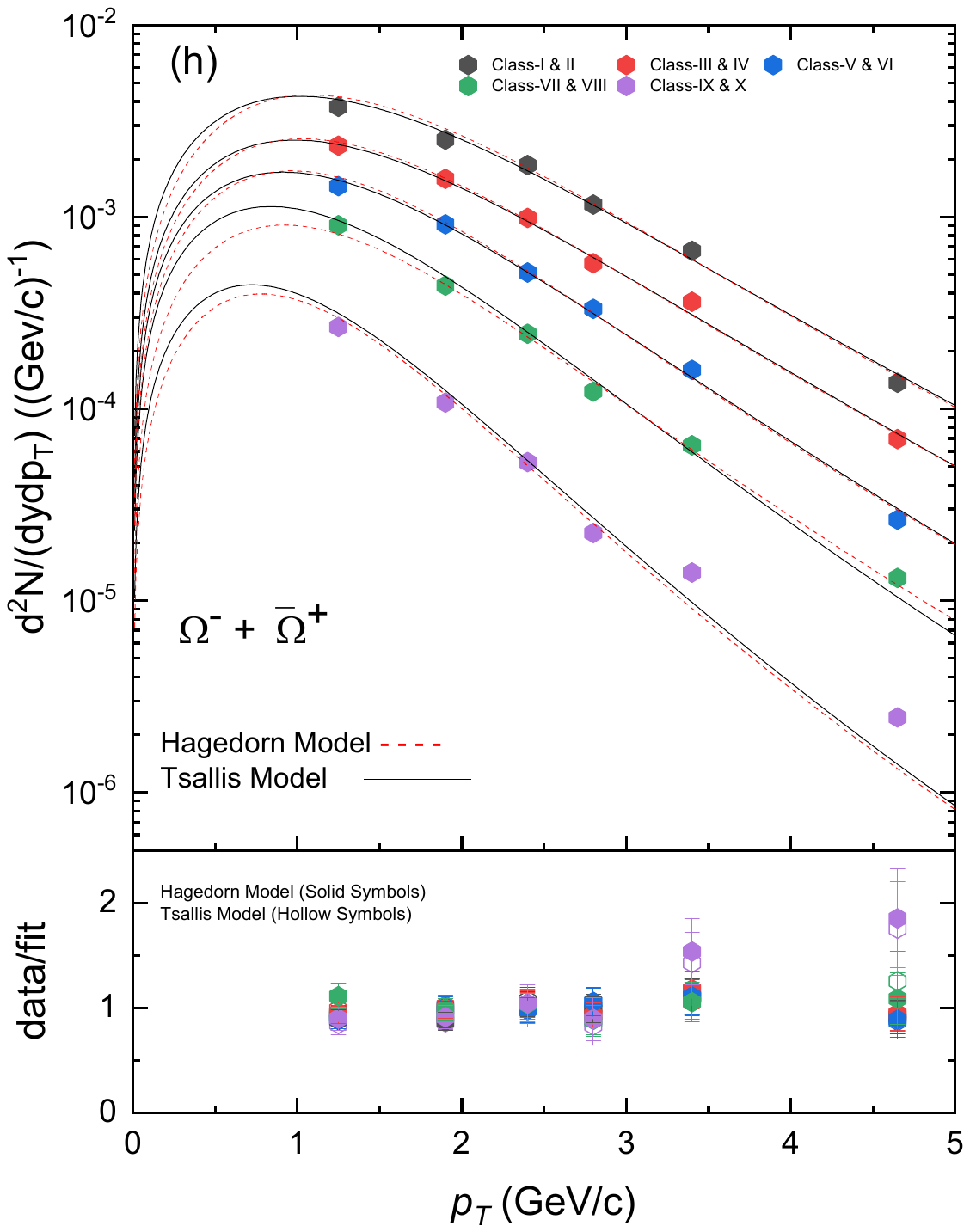}
\caption{The figure shows the $p_T$ spectra of the double differential yield of various light-flavored and strange hadrons at a center-of-mass energy of 7 TeV. These spectra were analyzed using both the Tsallis and Hagedorn models, with the results depicted using solid and dotted lines respectively. The experimental data points are illustrated using diverse colors and shapes. These analyses were conducted across multiple multiplicity classes.}
\label{fig:01}
\end{figure}

\begin{figure}[ht]
\centering
\includegraphics[width=9cm]{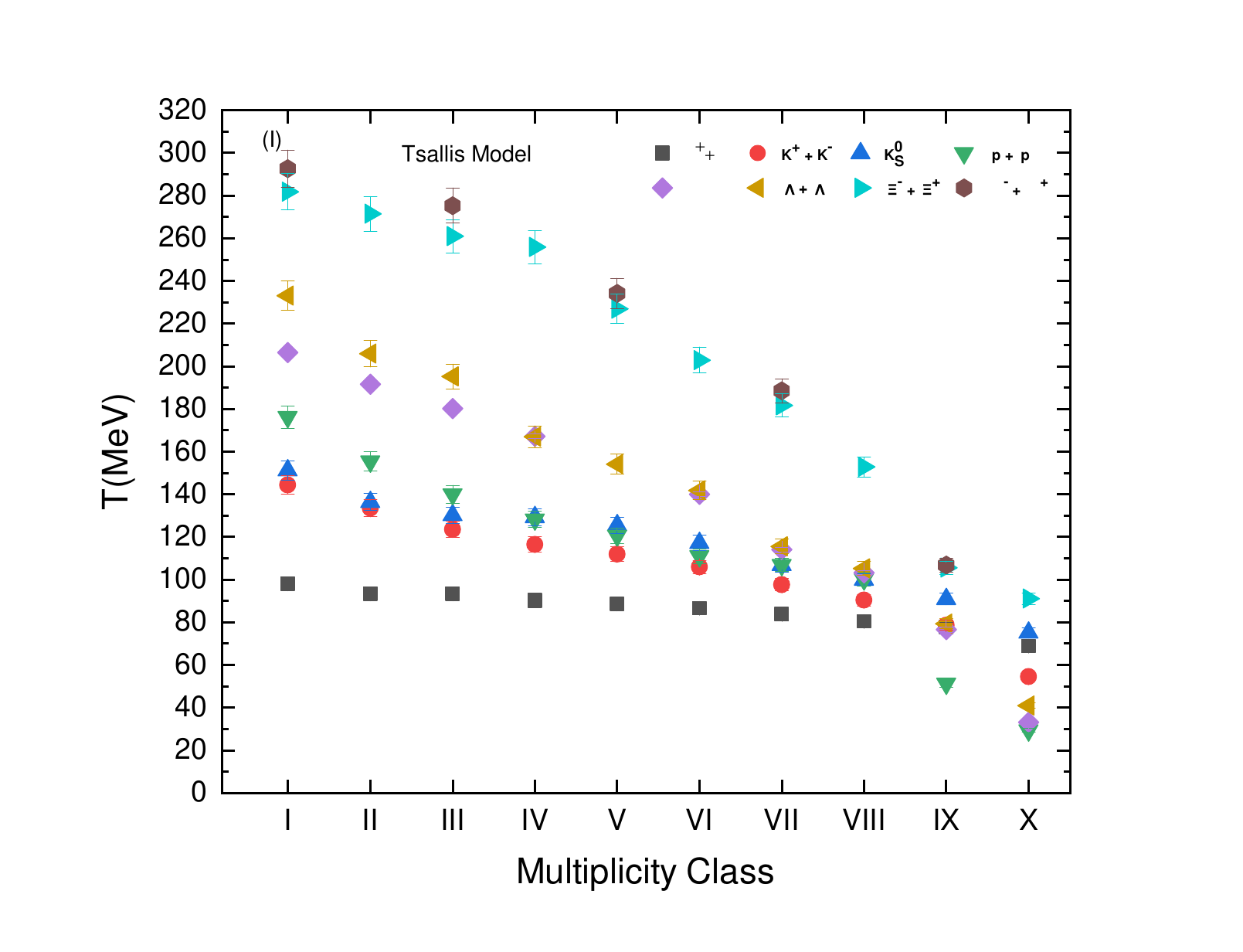}
\vspace{-0.7cm}
\hspace{-0.6cm}
\includegraphics[width=9cm]{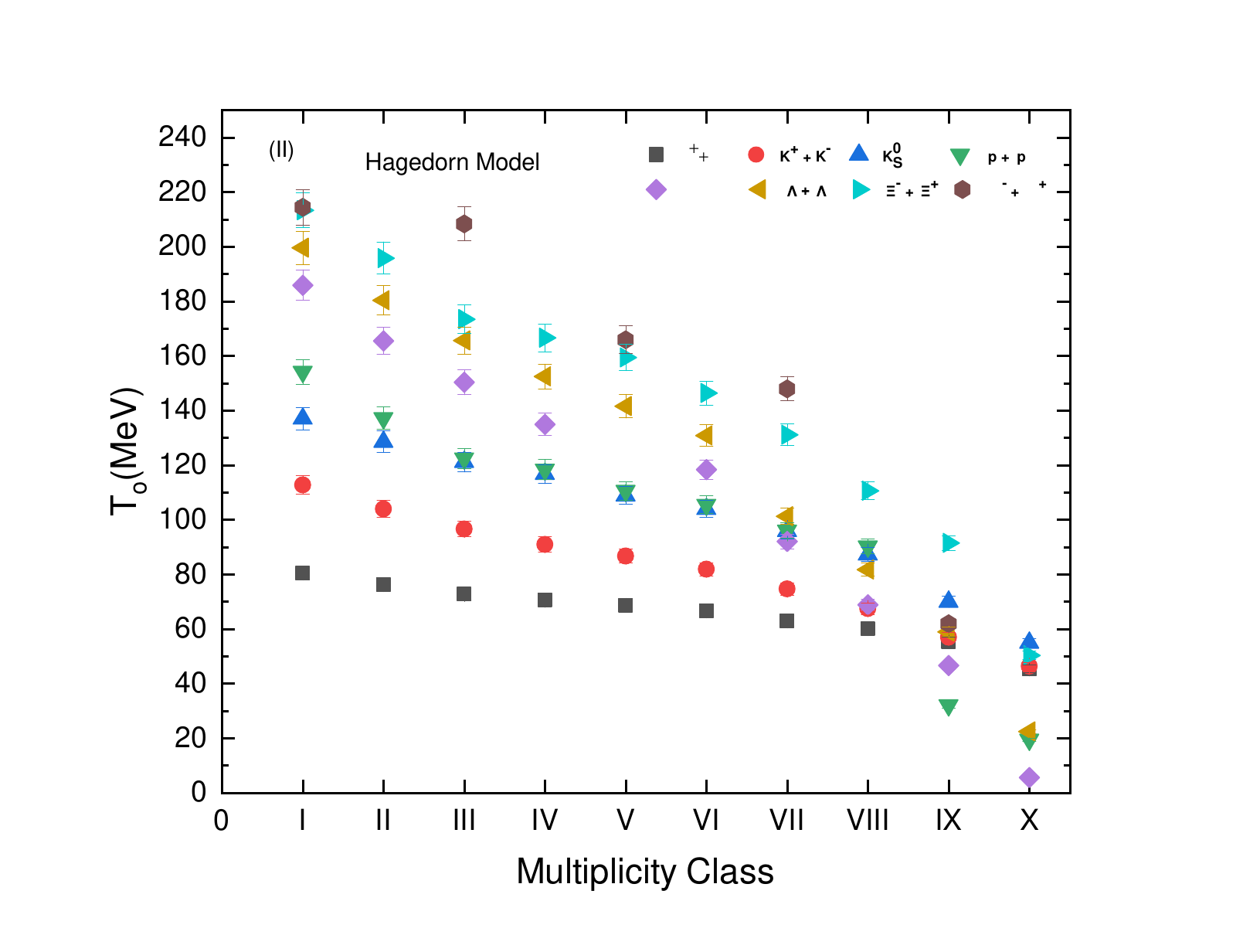}
\includegraphics[width=9cm]{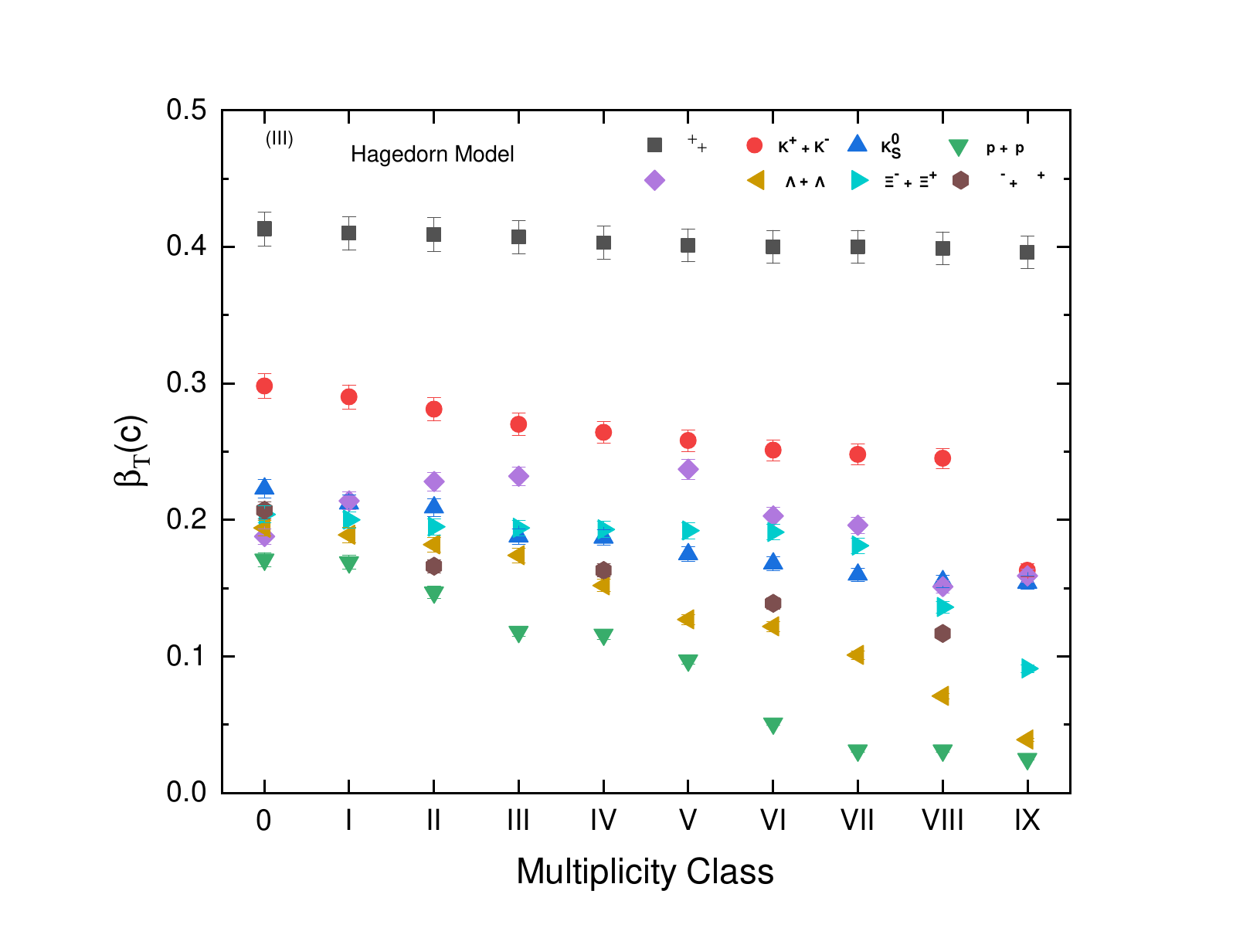}
\caption{Correlation plots showcasing diverse parameters obtained through the fitting of $p_T$ spectra for various light-flavored and strange hadrons, generated in proton-proton collisions at a center-of-mass energy of $\sqrt{s}$ = 7 TeV, across different multiplicity classes. The fitting process involves the utilization of both the Tsallis and Hagedorn models. These correlation plots offer insights into the relationships between the extracted parameters.}
\label{fig:02}
\end{figure}

\begin{figure}[ht]
\centering
\includegraphics[width=12cm]{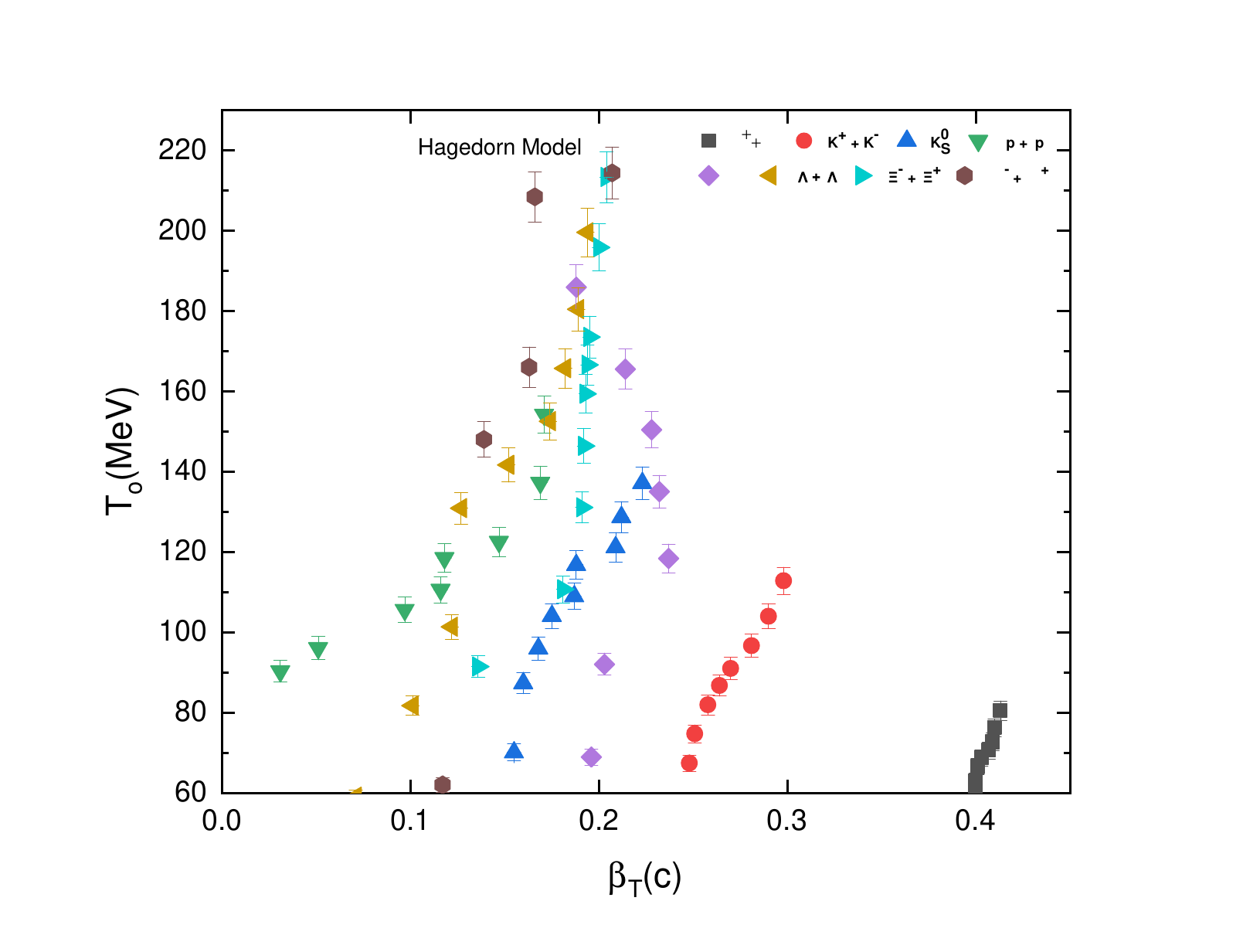}
\caption{A correlation plot depicting the relationship between the kinetic freeze-out temperature ($T_0$) and the transverse flow velocity ($\beta_T$) is presented. These parameters were extracted through the fitting procedure of the $p_T$ spectra of light-flavored and strange hadrons produced in proton-proton collisions at a center-of-mass energy of $\sqrt{s}$ = 7 TeV. The fitting was carried out employing the Hagedorn model, and the resulting correlation plot provides insights into the interplay between $T_0$ and $\beta_T$.}
\label{fig:03}
\end{figure}

To begin with the discussion of parameter trends, we introduce the plots in Figure \ref{fig:02}. As observed in Figure 2(i) and 2(ii), a noticeable decrease in both $T$ and $T_0$ is apparent when transitioning from multiplicity class I to class X. This trend can be attributed to the fact that in multiplicity class I, a substantial portion of the colliding systems significantly overlap, leading to a reduction in overlap as one progresses towards higher multiplicity classes. Consequently, this decrease in overlap results in a diminished energy transfer among the nucleons within the colliding systems. We noticed that $T$ and $T_0$ in Fig. 2(i) and 2(ii) of Fig. \ref{fig:02} follow a mass differential scenario which is compatible with \cite{m1} and our previous results \cite{m2,m3,m4}. Larger temperature for the heavier particles tends the heavier particles to decouple early from the system compared to the lighter particles. The reason behind the early freeze-out of the massive particles is that they have lower production rates due to higher energy requirements, therefore, they become less abundant and are more susceptible to freeze-out at higher temperatures compared to lighter particles. Besides, we noticed that the temperature (effective and kinetic freeze-out temperature) of the lighter particles is weakly dependent on multiplicity while its dependence on multiplicity becomes more significant as the particle mass increases, this result is consistent with the results reported by Khuntia et al., \cite{m1}. One potential explanation for this phenomenon could be as: Light particles have a lower kinetic freeze-out temperature, indicating weaker interactions with the surrounding medium. As a result, they are less responsive to variations in multiplicity. Conversely, heavier particles interact more strongly with the medium, rendering them more susceptible to changes in multiplicity. Apart from this, the values of transverse flow velocity in Fig. \ref{fig:02}(iii) is observed to be minimum at higher multiplicity classes and maximum at lower classes of multiplicity. In scenarios with lower multiplicity classes, more energy is transferred into the system, leading to a stronger pressure gradient in the collision zone and the creation of a highly compressed system. This compressed system holds considerable collision energy in potential form. Consequently, as the system begins to expand, it does so with a notably high transverse flow velocity. Conversely, higher multiplicity classes involve lesser energy transfer into the system, resulting in a less pronounced pressure gradient within the collision zone. This leads to a lower level of compression in the system, ultimately causing the expanding system to have a lower transverse flow velocity. Like $T_0$, $\beta_T$ also has a strong dependence for the heavier particles on multiplicity and weak dependence on multiplicity for the light particles. The correlation between $T_0$ and $\beta_T$ is depicted in Figure \ref{fig:03}, which reveals a positive correlation between the two parameters. The positive correlation exhibits high temperature and quick expansion of the system. The lower multiplicity class refers to the central collisions and the higher multiplicity class refers to the peripheral collisions. Hence the above result shows that the lower multiplicity class has a very high temperature and expands quickly. In some literature like ref. 14 the correlation between $T_0$ and $\beta_T$ is negative. Both the positive and negative correlations are correct and have their own explanations. The negative correlation shows the longer-lived fireball in lower multiplicity classes.
The positive correlation between kinetic freeze-out temperature and transverse flow velocity indicates that in high-energy collisions, particles with higher thermal energies also exhibit stronger collective motion. The higher multiplicity is associated with the higher energy transfer into the system, due to which the excitation function of the system increases which results in the hadronization of highly thermalized particles. On the other hand, due to the same higher energy transfer into the system at higher multiplicity, the system squeezes and then expands rapidly with greater $\beta_T$. Therefore, greater $\beta_T$ will always be accompanied by greater $T_0$ and vice versa.
Finally, the multiplicity parameter ($N_0$) displays a decline as the masses of produced particles increase, indicating a more prominent production of lighter particles in contrast to heavier ones. Additionally, $N_0$ demonstrates a diminishing trend as one progresses towards higher multiplicity classes. The connection between higher multiplicity classes and lower collision energies or centrality of particle collisions might explain the smaller values of $N_0$ in these scenarios.    

\begin{table}[!ht] 
\caption{\label{tab:1} Values of the free parameters for various identified particles including light-flavored and strange hadrons, derived from the Tsallis Model in hadronic collisions at a center-of-mass energy of 7 TeV.} 
    \centering
    \begin{tabular}{|c|c|c|c|c|c|c|c|}
    \hline
        \textbf{Energy} & \textbf{Particle} & \textbf{Multiplicity Class} & \textbf{T[MeV]} & \textbf{q} & \textbf{$N_0$} & \textbf{$\chi^2$} & \textbf{NDF}\\ \hline
         ~ & ~ & I & 97.900 ± 2.937 & 1.159 ± 0.035 & 295.400 ± 8.862 & 3.485 & 46 \\
        ~ & ~ & II & 93.200 ± 2.796 & 1.159 ± 0.035 & 232.400 ± 6.972 & 3.553 & 46 \\ 
        ~ & ~ & III & 93.200 ± 2.796 & 1.156 ± 0.035 & 189.000 ± 5.67 & 3.401 & 46 \\ 
        ~ & ~ & IV & 90.100 ± 2.703 & 1.156 ± 0.035 & 163.000 ± 4.89 & 2.943 & 46 \\ 
        7TeV & $\pi^+ + \pi^-$ & V & 88.500 ± 2.655 & 1.155 ± 0.035 & 144.500 ± 4.335 & 2.682 & 46 \\ 
        ~ & ~ & VI & 86.600 ± 2.598 & 1.154 ± 0.035 & 123.600 ± 3.708 & 2.304 & 46 \\ 
        ~ & ~ & VII & 83.600 ± 2.508 & 1.153 ± 0.035 & 98.800 ± 2.964 & 1.727 & 46 \\ 
        ~ & ~ & VIII & 80.400 ± 2.412 & 1.152 ± 0.035 & 81.400 ± 2.442 & 1.135 & 46 \\ 
        ~ & ~ & IX & 77.700 ± 2.331 & 1.149 ± 0.034 & 58.900 ± 1.767 & 0.597 & 46 \\ 
        ~ & ~ & X & 68.870 ± 2.066 & 1.142 ± 0.034 & 37.600 ± 1.128 & 0.425 & 46 \\ \hline
        ~ & ~ & I & 144.380 ± 4.331 & 1.145 ± 0.034 & 43.410 ± 1.302 & 18.52 & 41 \\ 
        ~ & ~ & II & 133.500 ± 4.005 & 1.146 ± 0.034 & 33.600 ± 1.008 & 13.517 & 41 \\ 
        ~ & ~ & III & 123.550 ± 3.707 & 1.149 ± 0.034 & 27.200 ± 0.816 & 12.965 & 41 \\ 
        ~ & ~ & IV & 116.520 ± 3.496 & 1.15 ± 0.035 & 23.160 ± 0.695 & 10.179 & 41 \\ 
        7TeV & $K^+ + K^-$ & V & 111.940 ± 3.358 & 1.151 ± 0.035 & 20.260 ± 0.608 & 7.344 & 41 \\ 
        ~ & ~ & VI & 105.920 ± 3.178 & 1.151 ± 0.035 & 16.950 ± 0.509 & 7.036 & 41 \\ 
        ~ & ~ & VII & 97.570 ± 2.927 & 1.152 ± 0.035 & 13.420 ± 0.403 & 6.408 & 41 \\ 
        ~ & ~ & VIII & 90.280 ± 2.708 & 1.153 ± 0.035 & 10.760 ± 0.323 & 3.455 & 41 \\ 
        ~ & ~ & IX & 78.540 ± 2.356 & 1.153 ± 0.035 & 7.740 ± 0.232 & 2.988 & 41 \\ 
        ~ & ~ & X & 54.520 ± 1.636 & 1.153 ± 0.035 & 4.590 ± 0.138 & 11.353 & 41 \\ \hline
        ~ & ~ & I & 151.180 ± 4.535 & 1.141 ± 0.034 & 21.310 ± 0.639 & 9.627 & 35 \\ 
        ~ & ~ & II & 136.390 ± 4.092 & 1.145 ± 0.034 & 16.600 ± 0.498 & 15.062 & 35 \\ 
        ~ & ~ & III & 130.120 ± 3.904 & 1.145 ± 0.034 & 13.380 ± 0.401 & 8.391 & 35 \\
        ~ & ~ & IV & 129.260 ± 3.878 & 1.142 ± 0.034 & 11.410 ± 0.342 & 15.115 & 35 \\ 
        7 TeV & $K_S^0$ & V & 125.460 ± 3.764 & 1.142 ± 0.034 & 9.940 ± 0.298 & 17.698 & 35 \\ 
        ~ & ~ & VI & 117.220 ± 3.517 & 1.144 ± 0.034 & 8.360 ± 0.251 & 17.015 & 35 \\ 
        ~ & ~ & VII & 106.880 ± 3.206 & 1.146 ± 0.034 & 6.650 ± 0.200 & 0.198 & 35 \\ 
        ~ & ~ & VIII & 99.920 ± 2.998 & 1.147 ± 0.034 & 5.280 ± 0.158 & 0.179 & 35 \\ 
        ~ & ~ & IX & 90.940 ± 2.728 & 1.146 ± 0.034 & 3.760 ± 0.113 & 0.165 & 35 \\ 
        ~ & ~ & X & 75.180 ± 2.255 & 1.143 ± 0.034 & 1.190 ± 0.036 & 22.741 & 35 \\ \hline
        ~ & ~ & I & 176.160 ± 5.285 & 1.114 ± 0.033 & 17.430 ± 0.523 & 23.575 & 39 \\ 
        ~ & ~ & II & 155.470 ± 4.664 & 1.117 ± 0.034 & 13.890 ± 0.417 & 38.499 & 39 \\
        ~ & ~ & III & 140.000 ± 4.200 & 1.118 ± 0.034 & 11.440 ± 0.343 & 31.564 & 39 \\
        ~ & ~ & IV & 128.250 ± 3.848 & 1.119 ± 0.034 & 9.890 ± 0.297 & 18.789 & 39 \\
        7 TeV & $p+\Bar{p}$ & V & 120.630 ± 3.619 & 1.12 ± 0.034 & 8.720 ± 0.262 & 28.602 & 39 \\
        ~ & ~ & VI & 110.990 ± 3.330 & 1.122 ± 0.034 & 7.320 ± 0.22 & 42.152 & 39 \\ 
        ~ & ~ & VII & 106.900 ± 3.207 & 1.117 ± 0.034 & 5.830 ± 0.175 & 39.765 & 39 \\ 
        ~ & ~ & VIII & 100.060 ± 3.002 & 1.115 ± 0.033 & 4.620 ± 0.139 & 50.804 & 39 \\ 
        ~ & ~ & IX & 51.190 ± 1.536 & 1.136 ± 0.034 & 3.440 ± 0.103 & 41.294 & 39 \\ 
        ~ & ~ & X & 29.420 ± 0.883 & 1.131 ± 0.034 & 1.780 ± 0.053 & 29.893 & 39 \\ \hline
        ~ & ~ & I & 206.420 ± 6.193 & 1.131 ± 0.034 & 2.650 ± 0.080 & 23.374 & 13 \\ 
        ~ & ~ & II & 191.520 ± 5.746 & 1.131 ± 0.034 & 2.090 ± 0.063 & 9.717 & 13 \\
        ~ & ~ & III & 180.210 ± 5.406 & 1.131 ± 0.034 & 1.730 ± 0.052 & 10.111 & 13 \\ 
        7TeV & $\phi$ & IV \& V & 167.220 ± 5.017 & 1.132 ± 0.034 & 1.380 ± 0.041 & 4.074 & 13 \\ 
        ~ & ~ & VI & 140.000 ± 4.200 & 1.137 ± 0.034 & 1.050 ± 0.032 & 9.938 & 13 \\ 
        ~ & ~ & VII & 114.090 ± 3.423 & 1.146 ± 0.034 & 0.820 ± 0.025 & 3.608 & 13 \\ 
        ~ & ~ & VIII & 103.150 ± 3.095 & 1.145 ± 0.034 & 0.630 ± 0.019 & 5.45 & 13 \\ 
        ~ & ~ & IX & 76.530 ± 2.296 & 1.151 ± 0.035 & 0.450 ± 0.014 & 6.28 & 13 \\ 
        ~ & ~ & X & 33.230 ± 0.997 & 1.152 ± 0.035 & 0.260 ± 0.008 & 3.07 & 13 \\ \hline
        \end{tabular}
\end{table}

        \begin{table}[!ht]
     To be continued.\\ \\
    \centering
    \begin{tabular}{|c|c|c|c|c|c|c|c|}
    \hline
        \textbf{Energy} & \textbf{Particle} & \textbf{Multiplicity Class} & \textbf{T[MeV]} & \textbf{q} & \textbf{$N_0$} & \textbf{$\chi^2$} & \textbf{NDF}\\ \hline
        ~ & ~ & I & 233.120 ± 6.994 & 1.091 ± 0.033 & 12.280 ± 0.368 & 8.107 & 13 \\ 
        ~ & ~ & II & 205.870 ± 6.176 & 1.095 ± 0.033 & 9.570 ± 0.287 & 5.738 & 13 \\ 
        ~ & ~ & III & 195.170 ± 5.855 & 1.094 ± 0.033 & 7.670 ± 0.230 & 5.822 & 13 \\ 
        ~ & ~ & IV & 166.970 ± 5.009 & 1.102 ± 0.033 & 6.550 ± 0.197 & 4.746 & 13 \\ 
        7 TeV & $\Lambda + \Bar{\Lambda}$ & V & 154.150 ± 4.625 & 1.104 ± 0.033 & 5.710 ± 0.171 & 4.152 & 13 \\ 
        ~ & ~ & VI & 141.830 ± 4.255 & 1.105 ± 0.033 & 4.710 ± 0.141 & 6.034 & 13 \\ 
        ~ & ~ & VII & 115.630 ± 3.469 & 1.112 ± 0.033 & 3.700 ± 0.111 & 4.753 & 13 \\ 
        ~ & ~ & VIII & 105.300 ± 3.159 & 1.112 ± 0.033 & 2.850 ± 0.086 & 6.541 & 13 \\ 
        ~ & ~ & IX & 79.380 ± 2.381 & 1.117 ± 0.034 & 1.920 ± 0.058 & 9.908 & 13 \\ 
        ~ & ~ & X & 40.920 ± 1.228 & 1.124 ± 0.034 & 0.850 ± 0.026 & 12.072 & 13 \\ \hline
        ~ & ~ & I & 281.770 ± 8.453 & 1.081 ± 0.032 & 1.550 ± 0.047 & 6.867 & 10 \\ 
        ~ & ~  & II & 271.550 ± 8.147 & 1.077 ± 0.032 & 1.130 ± 0.034 & 4.945 & 10 \\ 
        ~ & ~ & III & 260.930 ± 7.828 & 1.072 ± 0.032 & 0.900 ± 0.027 & 7.465 & 10 \\
        ~ & ~ & IV & 255.860 ± 7.676 & 1.072 ± 0.032 & 0.730 ± 0.022 & 6.738 & 10 \\ 
        7 TeV & $\Xi^- + \Bar{\Xi}^+$ & V & 226.900 ± 6.807 & 1.079 ± 0.032 & 0.620 ± 0.019 & 6.21 & 10 \\
        ~ & ~ & VI & 202.860 ± 6.086 & 1.084 ± 0.033 & 0.510 ± 0.015 & 7.544 & 10 \\
        ~ & ~ & VII & 181.730 ± 5.452 & 1.087 ± 0.033 & 0.380 ± 0.011 & 6.451 & 10 \\
        ~ & ~ & VIII & 152.780 ± 4.583 & 1.094 ± 0.033 & 0.290 ± 0.009 & 8.327 & 10 \\ 
        ~ & ~ & IX & 105.560 ± 3.167 & 1.108 ± 0.033 & 0.180 ± 0.005 & 5.9 & 10 \\ 
        ~ & ~ & X & 91.010 ± 2.730 & 1.102 ± 0.033 & 0.070 ± 0.002 & 11.781 & 10 \\ \hline
        ~ & ~ & I + II & 292.630 ± 8.779 & 1.079 ± 0.032 & 0.140 ± 0.004 & 2.63 & 3 \\ 
        ~ & ~ & III + IV & 275.340 ± 8.260 & 1.079 ± 0.032 & 0.080 ± 0.002 & 1.583 & 3 \\ 
        7 TeV & $\Omega^- + \Bar{\Omega}^+$ & V + VI & 234.160 ± 7.025 & 1.079 ± 0.032 & 0.080 ± 0.002 & 1.525 & 3 \\ 
        ~ & ~ & VII + VIII & 188.450 ± 5.654 & 1.080 ± 0.032 & 0.030 ± 0.001 & 3.411 & 3 \\ 
        ~ & ~ & IX + X & 106.850 ± 3.206 & 1.090 ± 0.033 & 0.010 ± 0.000 & 9.723 & 3 \\ \hline
    \end{tabular}
\end{table}

\begin{table}[!ht]
\caption{\label{tab:2}Values of the free parameters for distinct identified particles including light-flavored and strange hadrons, obtained through the application of the Hagedorn Model at a center-of-mass energy of 7 TeV.} 
    \centering
    \begin{tabular}{|c|c|c|c|c|c|c|c|c|}
    \hline
        \textbf{Energy} & \textbf{Particle} & \textbf{Multiplicity class} & \textbf{$T_0$ [MeV]} & \textbf{$\beta_T$ [c]} & \textbf{n} & \textbf{$N_0$} & \textbf{$\chi^2$} & \textbf{NDF} \\ \hline
        ~ & ~ & I & 80.500 ± 2.415 & 0.413 ± 0.012 & 6.350 ± 0.191 & 303.030 ± 9.091 & 2.943 & 46 \\ 
        ~ & ~ & II & 76.220 ± 2.287 & 0.410 ± 0.012 & 6.340 ± 0.190 & 237.600 ± 7.128 & 2.639 & 46  \\ 
        ~ & ~ & III & 72.720 ± 2.182 & 0.409 ± 0.012 & 6.340 ± 0.190 & 195.600 ± 5.868 & 2.417 & 46  \\ 
        ~ & ~ & IV & 70.560 ± 2.117 & 0.407 ± 0.012 & 6.340 ± 0.190 & 169.100 ± 5.073 & 2.087 & 46  \\ 
        7 Tev & $\pi^+ + \pi^-$ & V & 68.760 ± 2.063 & 0.403 ± 0.012 & 6.350 ± 0.191 & 150.300 ± 4.509 & 1.911 & 46 \\
        ~ & ~ & VI & 66.580 ± 1.997 & 0.401 ± 0.012 & 6.370 ± 0.191 & 128.200 ± 3.846 & 1.575 & 46 \\ 
        ~ & ~ & VII & 63.030 ± 1.891 & 0.400 ± 0.012 & 6.380 ± 0.191 & 104.120 ± 3.124 & 1.234 & 46 \\ 
        ~ & ~ & VIII & 60.190 ± 1.806 & 0.400 ± 0.012 & 6.400 ± 0.192 & 84.980 ± 2.549 & 0.868 & 46 \\
        ~ & ~ & IX & 55.340 ± 1.660 & 0.399 ± 0.012 & 6.470 ± 0.194 & 65.310 ± 1.959 & 0.711 & 46 \\ 
        ~ & ~ & X & 45.410 ± 1.362 & 0.396 ± 0.012 & 6.660 ± 0.200 & 42.040 ± 1.261 & 1.546 & 46 \\ \hline
        ~ & ~ & I & 112.800 ± 3.384 & 0.298 ± 0.009 & 6.660 ± 0.200 & 43.170 ± 1.295 & 5.873 & 41 \\ 
        ~ & ~ & II & 104.000 ± 3.120 & 0.290 ± 0.009 & 6.600 ± 0.198 & 33.340 ± 1.000 & 13.828 & 41 \\ 
        ~ & ~ & III & 96.680 ± 2.900 & 0.281 ± 0.008 & 6.530 ± 0.196 & 27.000 ± 0.810 & 14.799 & 41 \\ 
        ~ & ~ & IV & 91.020 ± 2.731 & 0.270 ± 0.008 & 6.450 ± 0.194 & 22.970 ± 0.689 & 13.013 & 41 \\ 
        7 Tev & $K^+ + K^-$ & V & 86.780 ± 2.603 & 0.264 ± 0.008 & 6.420 ± 0.193 & 20.060 ± 0.602 & 11.958 & 41 \\ 
        ~ & ~ & VI & 81.890 ± 2.457 & 0.258 ± 0.008 & 6.400 ± 0.192 & 16.780 ± 0.503 & 12.21 & 41 \\ 
        ~ & ~ & VII & 74.720 ± 2.242 & 0.251 ± 0.008 & 6.360 ± 0.191 & 13.290 ± 0.399 & 10.833 & 41 \\ 
        ~ & ~ & VIII & 67.450 ± 2.024 & 0.248 ± 0.007 & 6.320 ± 0.190 & 10.600 ± 0.318 & 8.61 & 41 \\ 
        ~ & ~ & IX & 56.990 ± 1.710 & 0.245 ± 0.007 & 6.330 ± 0.190 & 7.580 ± 0.227 & 9.838 & 41\\ 
        ~ & ~ & X & 46.430 ± 1.393 & 0.163 ± 0.005 & 6.420 ± 0.193 & 4.390 ± 0.132 & 5.421 & 41\\ \hline
        ~ & ~ & I & 137.120 ± 4.114 & 0.223 ± 0.007 & 6.960 ± 0.209 & 21.420 ± 0.643 & 20.389 & 35\\ 
        ~ & ~ & II & 128.670 ± 3.860 & 0.212 ± 0.006 & 6.890 ± 0.207 & 16.570 ± 0.497 & 0.312 & 35\\
        ~ & ~ & III & 121.200 ± 3.636 & 0.209 ± 0.006 & 6.870 ± 0.206 & 13.430 ± 0.403 & 0.233 & 35\\
        ~ & ~ & IV & 116.810 ± 3.504 & 0.188 ± 0.006 & 6.810 ± 0.204 & 11.470 ± 0.344 & 0.246 & 35\\
        7 TeV & $K_S^0$ & V & 109.010 ± 3.270 & 0.187 ± 0.006 & 6.710 ± 0.201 & 10.050 ± 0.302 & 0.248 & 35\\
        ~ & ~ & VI & 104.030 ± 3.121 & 0.175 ± 0.005 & 6.660 ± 0.200 & 8.390 ± 0.252 & 0.249 & 35 \\
        ~ & ~ & VII & 95.960 ± 2.879 & 0.168 ± 0.005 & 6.630 ± 0.199 & 6.690 ± 0.201 & 0.359 & 35\\
        ~ & ~ & VIII & 87.330 ± 2.620 & 0.160 ± 0.005 & 6.520 ± 0.196 & 5.320 ± 0.160 & 0.292 & 35 \\
        ~ & ~ & IX & 70.090 ± 2.103 & 0.155 ± 0.005 & 6.360 ± 0.191 & 3.850 ± 0.116 & 0.328 & 35 \\
        ~ & ~ & X & 54.970 ± 1.649 & 0.154 ± 0.005 & 6.550 ± 0.197 & 2.080 ± 0.062 & 23.985 & 35 \\ \hline
        ~ & ~  & I & 154.160 ± 4.625 & 0.171 ± 0.005 & 8.440 ± 0.253 & 17.280 ± 0.518 & 19.259 & 39 \\
        ~ & ~ & II & 137.260 ± 4.118 & 0.169 ± 0.005 & 8.370 ± 0.251 & 13.670 ± 0.410 & 33.515 & 39 \\
        ~ & ~ & III & 122.490 ± 3.675 & 0.147 ± 0.004 & 8.160 ± 0.245 & 11.290 ± 0.339 & 26.95 & 39 \\
        ~ & ~ & IV & 118.490 ± 3.555 & 0.118 ± 0.004 & 8.160 ± 0.245 & 9.750 ± 0.293 & 16.763 & 39\\
        7 TeV & $p + \Bar{p}$ & V & 110.590 ± 3.318 & 0.116 ± 0.003 & 8.100 ± 0.243 & 8.600 ± 0.258 & 25.739 & 39\\
        ~ & ~ & VI & 105.590 ± 3.168 & 0.097 ± 0.003 & 8.050 ± 0.242 & 7.190 ± 0.216 & 39.377 & 39\\
        ~ & ~ & VII & 96.090 ± 2.883 & 0.051 ± 0.002 & 7.920 ± 0.238 & 5.880 ± 0.176 & 21.055 & 39 \\
        ~ & ~ & VIII & 90.290 ± 2.709 & 0.031 ± 0.001 & 7.920 ± 0.238 & 4.730 ± 0.142 & 16.887 & 39 \\
        ~ & ~ & IX & 32.090 ± 0.963 & 0.031 ± 0.001 & 6.900 ± 0.207 & 3.330 ± 0.100 & 43.803 & 39 \\
        ~ & ~ & X & 19.520 ± 0.586 & 0.025 ± 0.001 & 7.420 ± 0.223 & 1.810 ± 0.054 & 18.87 & 39 \\ \hline
        ~ & ~ & I & 185.940 ± 5.578 & 0.188 ± 0.006 & 7.550 ± 0.227 & 2.63 ± 0.079 & 21.657 & 13 \\
        ~ & ~ & II & 165.530 ± 4.966 & 0.214 ± 0.006 & 7.620 ± 0.229 & 2.05 ± 0.062 & 6.609 & 13 \\
        ~ & ~ & III & 150.430 ± 4.513 & 0.228 ± 0.007 & 7.580 ± 0.227 & 1.680 ± 0.050 & 9.056 & 13 \\
        ~ & ~ & IV \& V & 135.000 ± 4.050 & 0.232 ± 0.007 & 7.540 ± 0.226 & 1.330 ± 0.040 & 4.309 & 13 \\
        7 TeV & $\phi$ & VI & 118.400 ± 3.552 & 0.237 ± 0.007 & 7.510 ± 0.225 & 1.000 ± 0.030 & 7.038 & 13 \\
        ~ & ~ & VII & 92.070 ± 2.762 & 0.203 ± 0.006 & 6.880 ± 0.206 & 0.780 ± 0.023 & 4.749 & 13 \\
        ~ & ~ & VIII & 68.940 ± 2.068 & 0.196 ± 0.006 & 6.670 ± 0.200 & 0.610 ± 0.018 & 5.4 & 13 \\
        ~ & ~ & IX & 46.620 ± 1.399 & 0.151 ± 0.005 & 6.340 ± 0.190 & 0.430 ± 0.013 & 7.85 & 13 \\
        ~ & ~ & X & 5.690 ± 0.171 & 0.159 ± 0.005 & 6.340 ± 0.190 & 0.240 ± 0.007 & 6.316 & 13 \\ \hline
         \end{tabular}
\end{table}

\begin{table}[!ht]
 To be continued.\\ \\
    \centering
    \begin{tabular}{|c|c|c|c|c|c|c|c|c|}
    \hline
        \textbf{Energy} & \textbf{Particle} & \textbf{Multiplicity class} & \textbf{$T_0$ [MeV]} & \textbf{$\beta_T$ [c]} & \textbf{n} & \textbf{$N_0$} & \textbf{$\chi^2$} & \textbf{NDF} \\ \hline
        ~ & ~ & I & 199.570 ± 5.987 & 0.194 ± 0.006 & 10.500 ± 0.315 & 12.040 ± 0.361 & 5.608 & 13 \\
        ~ & ~ & II & 180.410 ± 5.412 & 0.189 ± 0.006 & 10.390 ± 0.312 & 9.360 ± 0.281 & 5.686 & 13\\
        ~ & ~ & III & 165.710 ± 4.971 & 0.182 ± 0.005 & 10.180 ± 0.305 & 7.500 ± 0.225 & 7.912 & 13 \\
        ~ & ~ & IV & 152.510 ± 4.575 & 0.174 ± 0.005 & 9.950 ± 0.299 & 6.330 ± 0.190 & 10.786 & 13\\
        7 TeV & $\Lambda + \Bar{\Lambda}$ & V & 141.690 ± 4.251 & 0.152 ± 0.005 & 9.620 ± 0.289 & 5.510 ± 0.165 & 8.709 & 13\\
        ~ & ~ & VI & 130.930 ± 3.928 & 0.127 ± 0.004 & 9.390 ± 0.282 & 4.580 ± 0.137 & 9.261 & 13\\
        ~ & ~ & VII & 101.340 ± 3.040 & 0.122 ± 0.004 & 8.720 ± 0.262 & 3.590 ± 0.108 & 8.024 & 13 \\
        ~ & ~ & VIII & 81.780 ± 2.453 & 0.101 ± 0.003 & 8.310 ± 0.249 & 2.810 ± 0.084 & 6.987 & 13\\
        ~ & ~ & IX & 59.000 ± 1.770 & 0.071 ± 0.002 & 7.960 ± 0.239 & 1.910 ± 0.057 & 6.421 & 13 \\
        ~ & ~ & X & 22.500 ± 0.675 & 0.039 ± 0.001 & 7.550 ± 0.227 & 0.860 ± 0.026 & 6.798 & 13  \\ \hline
        ~ & ~ & I & 213.330 ± 6.400 & 0.204 ± 0.006 & 10.540 ± 0.316 & 1.530 ± 0.046 & 8.642 & 10 \\
        ~ & ~ & II & 195.830 ± 5.875 & 0.200 ± 0.006 & 10.480 ± 0.314 & 1.110 ± 0.033 & 3.58 & 10 \\
        ~ & ~ & III & 173.500 ± 5.205 & 0.195 ± 0.006 & 10.470 ± 0.314 & 0.900 ± 0.027 & 4.177 & 10 \\
        ~ & ~ & IV & 166.600 ± 4.998 & 0.194 ± 0.006 & 10.440 ± 0.313 & 0.720 ± 0.022 & 2.182 & 10 \\
        7 TeV & $\Xi^- + \Bar{\Xi}^+$ & V & 159.420 ± 4.783 & 0.193 ± 0.006 & 10.410 ± 0.312 & 0.590 ± 0.018 & 4.173 & 10\\
        ~ & ~ & VI & 146.410 ± 4.392 & 0.192 ± 0.006 & 10.340 ± 0.310 & 0.480 ± 0.014 & 6.928 & 10 \\
        ~ & ~ & VII & 131.140 ± 3.934 & 0.191 ± 0.006 & 10.230 ± 0.307 & 0.340 ± 0.010 & 8.033 & 10 \\
        ~ & ~ & VIII & 110.690 ± 3.321 & 0.181 ± 0.005 & 9.850 ± 0.296 & 0.270 ± 0.008 & 10.504 & 10 \\
        ~ & ~ & IX & 91.480 ± 2.744 & 0.136 ± 0.004 & 9.140 ± 0.274 & 0.160 ± 0.005 & 12.346 & 10 \\
        ~ & ~ & X & 50.430 ± 1.513 & 0.091 ± 0.003 & 8.610 ± 0.258 & 0.070 ± 0.002 & 7.703 & 10 \\ \hline
        ~ & ~ & I + II & 214.420 ± 6.433 & 0.207 ± 0.006 & 11.270 ± 0.338 & 0.140 ± 0.004 & 4.202 & 3 \\
        ~ & ~ & III + IV & 208.420 ± 6.253 & 0.166 ± 0.005 & 10.930 ± 0.328 & 0.080 ± 0.002 & 1.732 & 3 \\
        7 TeV & $\Omega^- + \Bar{\Omega}^+$ & V + VI & 166.020 ± 4.981 & 0.163 ± 0.005 & 10.920 ± 0.328 & 0.050 ± 0.002 & 2.274 & 3 \\
        ~ & ~ & VII + VIII & 148.020 ± 4.441 & 0.139 ± 0.004 & 10.560 ± 0.317 & 0.025 ± 0.001 & 1.771 & 3 \\
        ~ & ~ & IX + X & 62.020 ± 1.861 & 0.117 ± 0.004 & 9.990 ± 0.300 & 0.009 ± 0.000 & 8.092 & 3 \\ \hline
    \end{tabular}
\end{table}

\section*{Conclusion}
The analysis of transverse momentum spectra for identified particles, encompassing light-flavored and strange hadrons, utilized the Tsallis and Hagedorn models. It is found that both models demonstrate a satisfactory fit with the experimental data. We have extracted the effective temperature ($T$), kinetic freeze-out temperature ($T_0$), and transverse flow velocity ($\beta_T$). These parameters exhibit an increase as we move towards lower multiplicity classes, driven by the higher energy transfer in such cases. It is important to note that both $T_{eff}$ and $T_0$ demonstrate an upward trend with the rising masses of particle species, thereby confirming the existence of a multi-freeze-out scenario. In this scenario, lighter particles experience freeze-out later than the heavier particles.

The normalization constant, or the multiplicity parameter, shows a direct correlation with collision event multiplicity, underscoring reduced particle production in higher multiplicity classes. Conversely, this constant exhibits an inverse relationship with the masses of the generated particles, indicating diminished production of heavier hadrons compared to lighter ones. Furthermore, a weak dependence of temperature (effective and kinetic freeze-out) on multiplicity is observed for lighter particles, while heavier particles exhibit a strong temperature dependence. One possible explanation for this trend is as follows: Light particles tend to possess a lower kinetic freeze-out temperature, indicating less interaction with the surrounding medium. Consequently, they exhibit reduced sensitivity to shifts in multiplicity. Conversely, heavier particles engage in stronger interactions with the medium, rendering them more responsive to changes in multiplicity.

Moreover, our study reveals a positive correlation between transverse flow velocity and kinetic freeze-out temperature. This positive correlation suggests that there is a higher degree of excitation in the lower multiplicity classes (higher multiplicity events) which corresponds to higher temperature and quick expansion.

\section*{Acknowledgements}
 The present research work was funded by Princess Nourah bint Abdulrahman University Researchers Supporting Project number (PNURSP2024R106), Princess Nourah bint Abdulrahman University, Riyadh, Saudi Arabia. We would like to express our gratitude for the support received from Abdul Wali Khan University Mardan, Pakistan; Hubei University of Automotive Technology Doctoral Research Fund under Grant Number BK202313; University of Guyana; University of Tabuk, Saudi Arabia; and Qassim University, Saudi Arabia that have contributed to creating a conducive research environment.
\section*{CRediT authorship contribution statement}
M.Ajaz: Conceptualization, Methodology, Writing original draft. M. Shehzad: Formal analysis, Visualization. M. Waqas: Formal analysis, Visualization. H.I. Alrebdi: Methodology, Writing review \& editing, Funding acquisition, Visualization. A. Jagnandan: Software, Methodology. M.A. Ahmad: Software, Methodology. S. Jagnandan: Software, Formal analysis. M. Badshah: Methodology, Writing original draft. J.H. Baker: Methodology, Investigation, Writing review \& editing. A.M. Quraishi: Supervision, Methodology.
\section*{Declaration of Competing Interest} 
The authors declare that there are no known financial interests or personal relationships that could have potentially influenced the findings presented in this paper.
\section*{Data Availability}
The data utilized in this research is either provided within the manuscript itself or appropriately referenced at relevant points.
\bibliography{sample}
\end{document}